\documentclass[11pt]{article}
\usepackage{times}
\usepackage{pdfpages}
\usepackage{setspace}
\usepackage{sectsty}
\usepackage{enumitem}

\sectionfont{\fontsize{11}{15}\selectfont}
\subsectionfont{\fontsize{11}{15}\selectfont}

\usepackage{wrapfig}
\usepackage{lipsum}
\usepackage{graphicx}

\usepackage{array}
\usepackage{subfigure}
\usepackage{booktabs}
\usepackage{multicol}
\usepackage{multirow}
\usepackage{threeparttable}
\usepackage[small, bf]{caption}
\usepackage[tableposition=top]{caption}
\usepackage{booktabs}
\usepackage{siunitx}

\usepackage{pgfgantt}

\usepackage{wrapfig}
\usepackage{times}
\usepackage{url}
\usepackage{soul}
\usepackage{color}
\begingroup
\makeatletter
\g@addto@macro{\UrlSpecials}{%
  \endlinechar=13 \catcode\endlinechar=12
  \do\%{\Url@percent}\do\^^M{\break}}
 \gdef\Url@percent{\@ifnextchar^^M{\@gobble}{\mathbin{\mathchar`\%}}}%
\endgroup %
\usepackage{array}
\usepackage{subfigure}

\usepackage{array}
\newcolumntype{L}[1]{>{\raggedright\let\newline\\\arraybackslash\hspace{0pt}}m{#1}}

\usepackage[headheight=14pt,tmargin=1in,bmargin=1in,lmargin=1in,rmargin=1in]{geometry}

\usepackage{fancyhdr}
\usepackage{datetime}
\usepackage{indentfirst}
\usepackage[sort&compress, square, comma, numbers]{natbib}

\usepackage[compact]{titlesec}

\usepackage{titlesec}
\titlespacing*{\section}{0pt}{0.1\baselineskip}{0.1\baselineskip}
\titlespacing*{\subsection}{0pt}{0.2\baselineskip}{0.1\baselineskip}
\titlespacing*{\subsubsection}{0pt}{0.2\baselineskip}{0.1\baselineskip}
\titlespacing*{\paragraph}{0pt}{0.2\baselineskip}{0.2\baselineskip}

\usepackage{blindtext}

\usepackage{threeparttable}
\usepackage{soul}
\usepackage{booktabs}
\usepackage{multirow}

\usepackage{makecell} 

\begin{document}
\thispagestyle{empty}
\begin{center}
{\bf \Large Categorizing Social Media Screenshots for Identifying\\Author Misattribution}\\

\vspace{5mm}
Ashlyn M. Farris\\
Mathematics Department \\
Harding University \\
afarris4@harding.edu\\
\hfill \break
Michael L. Nelson\\
Department of Computer Science \\
Virginia Modeling, Analysis and Simulation Center\\
Old Dominion University\\
mln@cs.odu.edu \\
\end{center}

\begin{abstract}
Mis/disinformation is a common and dangerous occurrence on social media. Misattribution is a form of mis/disinformation that deals with a false claim of authorship, which means a user is claiming someone said (posted) something they never did. We discuss the difference between misinformation and disinformation and how screenshots are used to spread author misattribution on social media platforms. It is important to be able to find the original post of a screenshot to determine if the screenshot is being correctly attributed. To do this we have built several tools to aid in automating this search process. The first is a Python script that aims to categorize Twitter posts based on their structure, extract the metadata from a screenshot, and use this data to group all the posts within a screenshot together. We tested this process on 75 Twitter posts containing screenshots collected by hand to determine how well the script extracted metadata and grouped the individual posts, $F_1=0.80$. The second is a series of scrapers being used to collect a dataset that can train and test a model to differentiate between various social media platforms. We collected 16,620 screenshots have been collected from Facebook, Instagram, Truth Social, and Twitter. Screenshots were taken by the scrapers of the web version and mobile version of each platform in both light and dark mode.
\end{abstract}
\setcounter{page}{1}
\section{Introduction}
When talking about false information we generally refer to two categories, misinformation and disinformation. The main difference between these two categories is the intent behind them. In this section, we will discuss misattribution, a form of mis/disinformation, and how screenshots are used both as tools to extend the functionality of a social media platform and spread author misattribution. We will also discuss the common methods used to locate the original post of a screenshot.

\subsection{Misinformation and Disinformation}
\begin{figure}
\begin{center}
\includegraphics[scale=0.5]{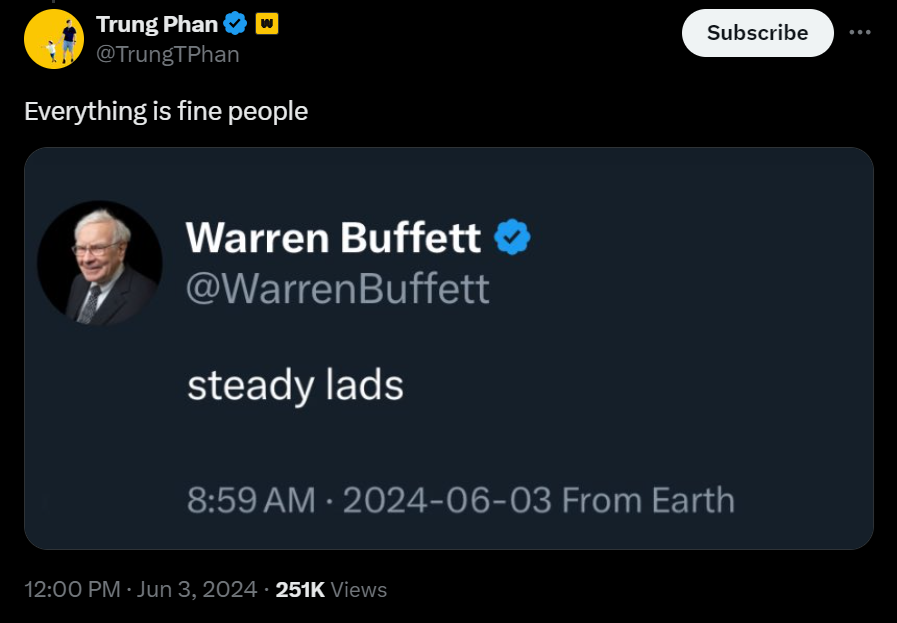}%
\captionof{figure}{Screenshot of fabricated post, attributed to @WarrenBuffett, shared on Twitter.\protect}\label{labelname}%
\label{fig:WarrenBuffett}
\end{center}
\end{figure}

The two main terms we use when referring to false information are misinformation and disinformation. The defining difference between these two categories is the intent with which the information is shared. Misinformation is false information that is shared without the intent to mislead its audience. In other words, a person shares information they believe without knowing it is untrue. Disinformation, however, is information that is shared with the intent to mislead its audience \cite{Apa-Misinformation/Disinformation}. The focus of this report is mis/disinformation in the form of misattribution. When identifying author misattribution, we are not concerned with the content of information but if the information is attributed to the correct person. If we take the tweet posted by @TrungTPhan (Figure \ref{fig:WarrenBuffett}) as an example,\footnotemark\footnotetext{\url{https://x.com/TrungTPhan/status/1797659696712826969}} we can see that a screenshot of a post supposedly made by @WarrenBuffett is being shared. We are not concerned if the information in @WarrenBuffett’s post is true, but if that post was created by @WarrenBuffett. This screenshot, though intended to be humorous, 
 is an example of a fabricated screenshot and is not correctly attributed.

\subsection{Screenshots}
\begin{figure}
\begin{center}
\includegraphics[scale=0.5]{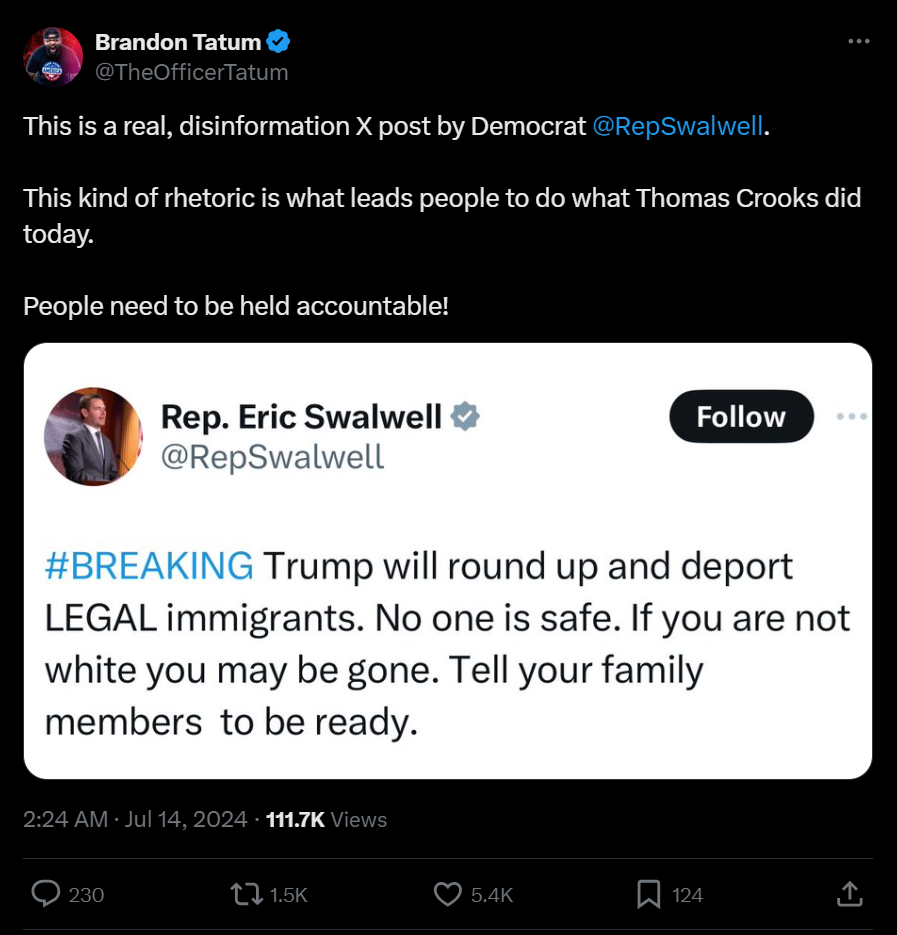}%
\captionof{figure}{Screenshot of a Twitter post, attributed to @RepSwalwell, shared on Twitter by @TheOfficerTatum, shared to demonstrate perceived disinformation.\protect}\label{labelname}%
\label{fig:denyEngagement}
\end{center}
\end{figure}

\begin{figure}
\begin{center}
\includegraphics[scale=0.5]{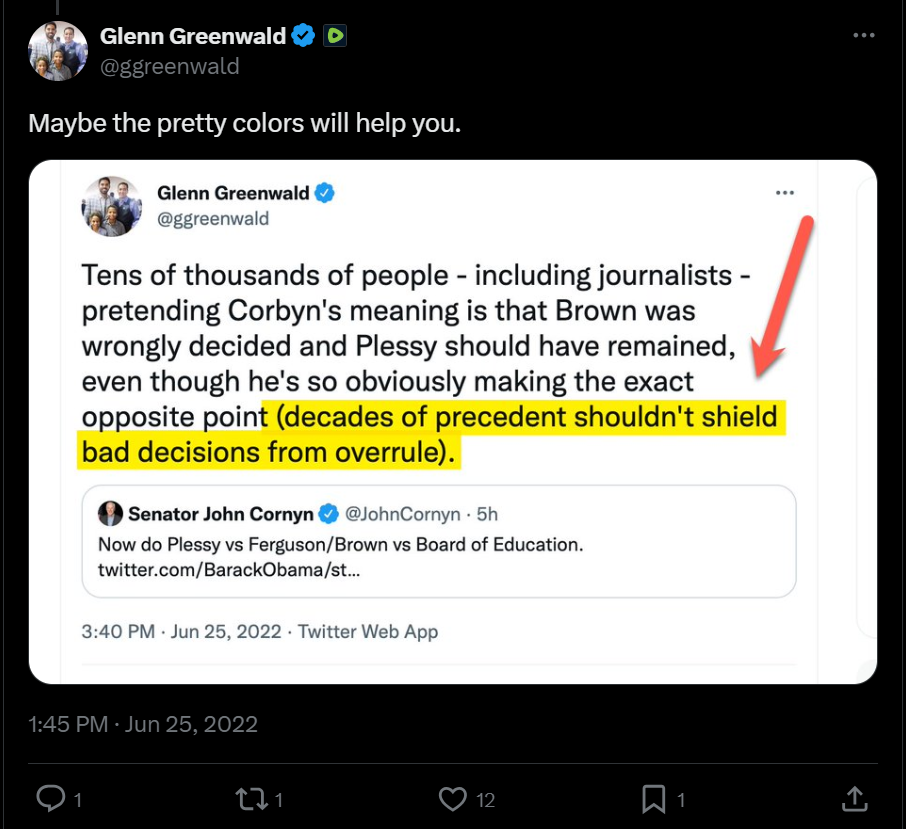}%
\captionof{figure}{Screenshot of a post with yellow highlights and a red arrow added by @ggreenwald.}\label{labelname}%
\label{fig:annotatedScreenshot}
\end{center}
\end{figure}

\begin{figure}
\begin{center}
\includegraphics[scale=0.2]{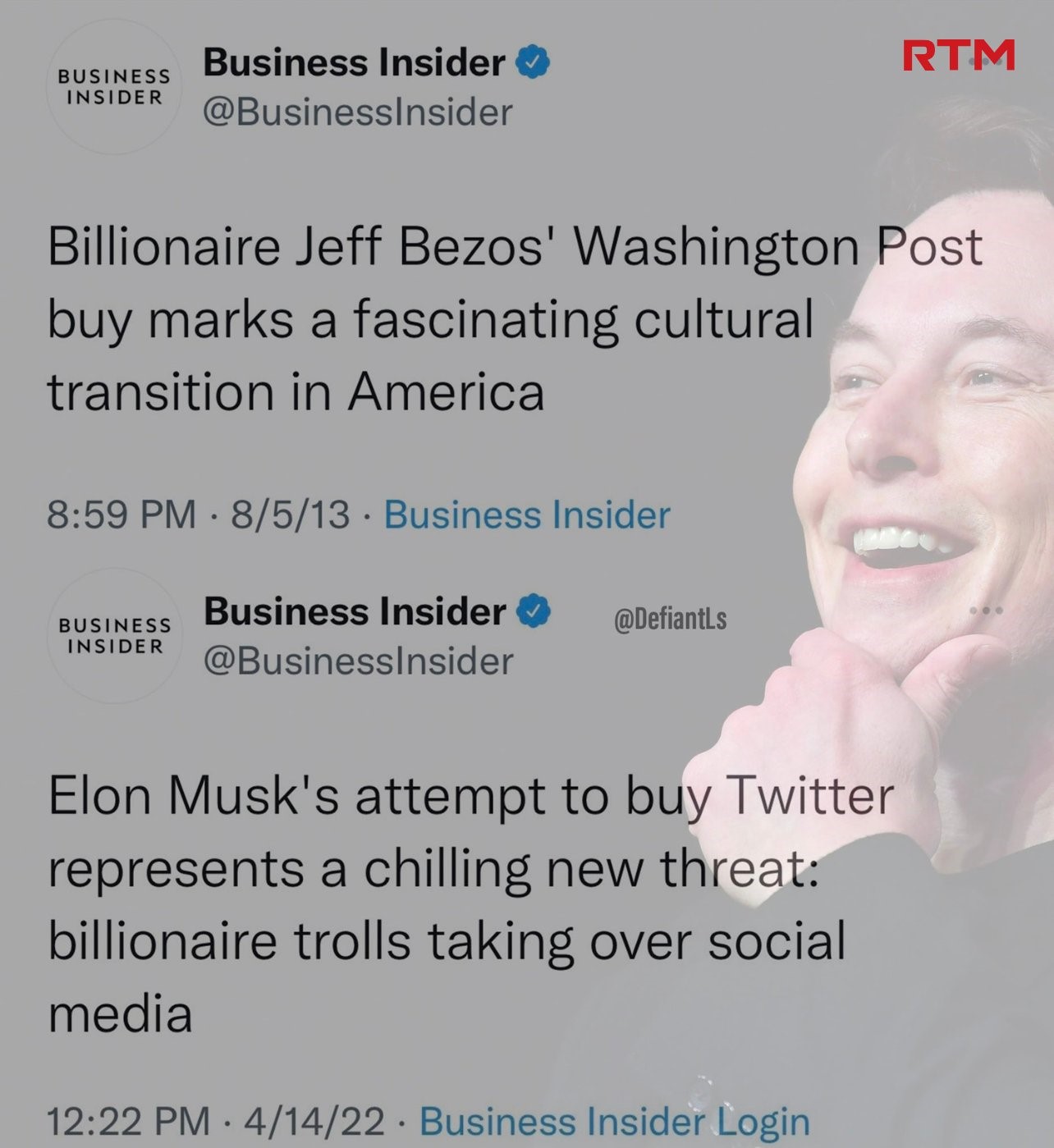}%
\captionof{figure}{Two posts shared on Twitter concatenated together to show perceived polar opinions posted by the same user.}\label{labelname}%
\label{fig:concatenatedScreenshot}
\end{center}
\end{figure}

\begin{figure}
\begin{center}
\includegraphics[scale=0.5]{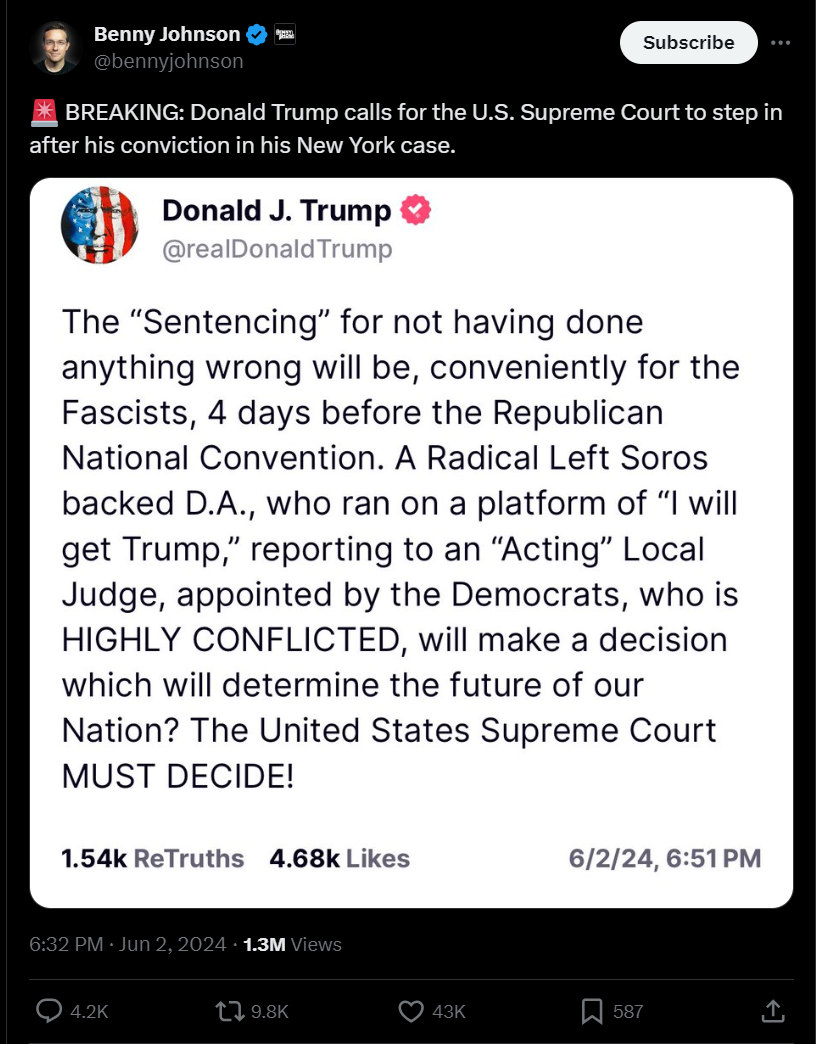}%
\captionof{figure}{Screenshot of a Truth Social post shared on Twitter demonstrating cross-platform sharing.\protect\footnotemark}\label{labelname}%
\label{fig:crossPlatform}
\end{center}
\footnotetext{\url{https://x.com/bennyjohnson/status/1797411023881654627}}
\end{figure}

\begin{figure}
\begin{center}
\includegraphics[scale=0.5]{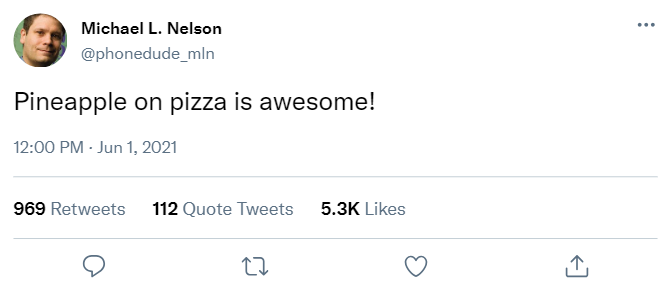}%
\captionof{figure}{Example of a fake tweet generated using Tweetgen (whether or not pineapple belongs on pizza is an ongoing controversy in our research group).}%
\label{fig:pineappleTweet}
\end{center}
\end{figure}

One way that misattribution commonly occurs is through screenshots shared on social media (Figure \ref{fig:WarrenBuffett}). In this context, a screenshot is an image of the contents of a social media post and then shared as an image, instead of via the platform's own sharing idioms (e.g., Quote Tweeting).  If a screenshot is taken of a social media post and then shared there is no direct link back to the original post to verify if the image is attributed correctly. Though screenshots can be fabricated and used to spread disinformation, they are often used to extend a platform's functionality. These screenshots can be used to deter engagement the original post would receive. This helps users to share information without allowing a post they disagree with or is perceived as harmful to gain traction and be spread to more people (Figure \ref{fig:denyEngagement}). For example, this post\footnotemark\footnotetext{\url{https://x.com/TheOfficerTatum/status/1812372624347222391}} authored by @TheOfficerTatum is shared claiming the post authored by @RepSwalwell contains disinformation. By sharing this screenshot @TheOfficerTatum can discuss this perceived disinformation without increasing the comments, likes, and reposts of the original tweet written by @RepSwalwell. Screenshots can also be edited to add emphasis or highlight important parts of a post (Figure \ref{fig:annotatedScreenshot}), such as the yellow highlights added to this screenshot,\protect\footnotemark\footnotetext{\url{https://x.com/ggreenwald/status/1540768229651410944}} or pasted together to show consensus or disagreement (Figure \ref{fig:concatenatedScreenshot}). This image\protect\footnotemark\footnotetext{\url{https://x.com/DefiantLs/status/1794138809254449167}} demonstrates two posts pasted together to create a concatenated screenshot. Additionally, screenshots can be used to share posts across different platforms. Many social media platforms do not have a way to share a post directly across several platforms, screenshots are used to do this (Figure \ref{fig:crossPlatform}). It is quite simple to create a fake social media post using online tools such as {TweetGen\protect\footnotemark}\label{labelname}\footnotetext{\url{https://www.tweetgen.com/create/tweet.html}} (Figure \ref{fig:pineappleTweet}), {GenerateStatus,\protect\footnotemark}\label{labelname}\footnotetext{\url{https://generatestatus.com/}} or developer tools. These images can then be posted on a social media platform and are indistinguishable from a screenshot of a real post. 

\subsection{Searching for the Alleged Original Post in the Screenshot}
\begin{figure}
\begin{center}
\includegraphics[scale=0.5]{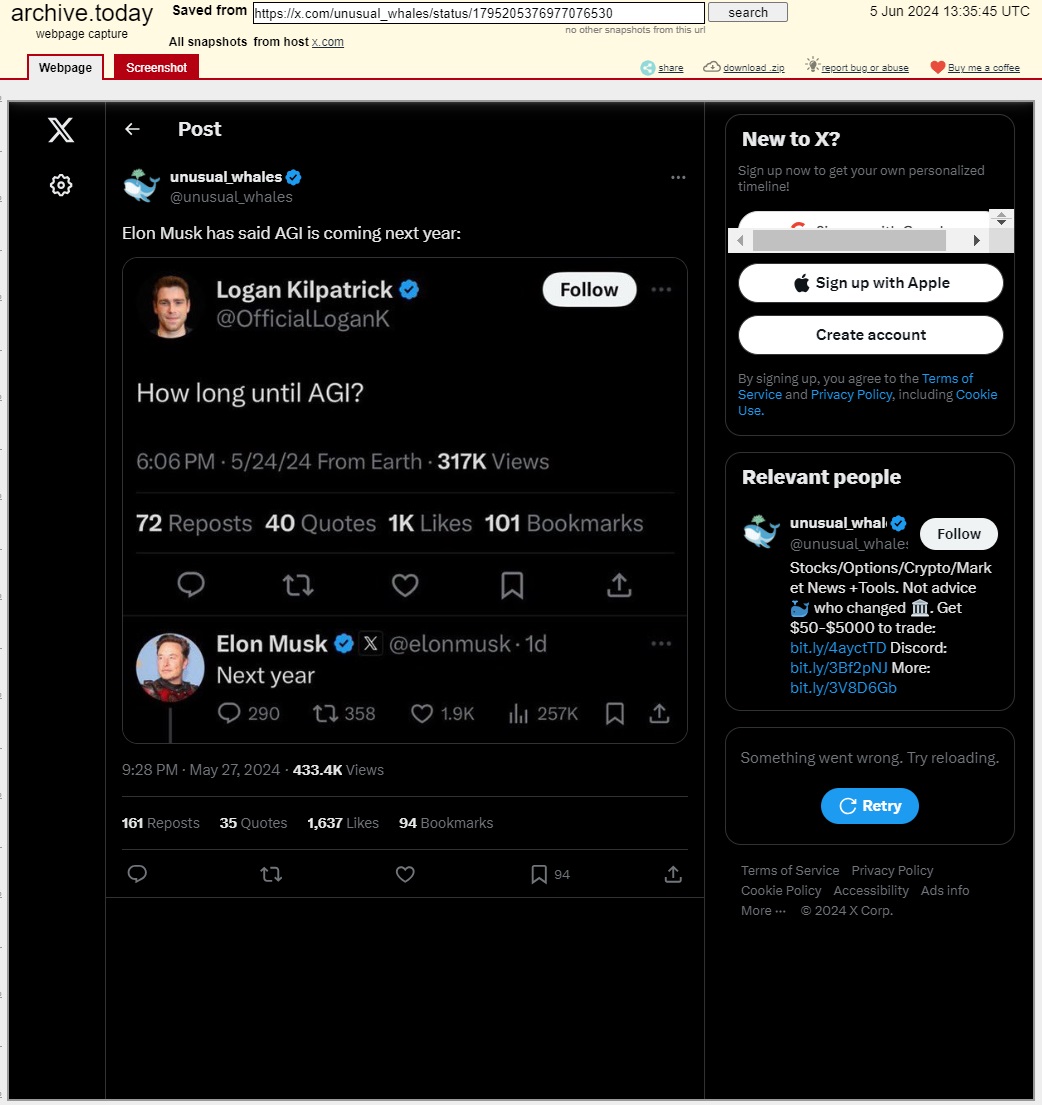}%
\captionof{figure}{Twitter post of a screenshot saved on web archive Archive.Today.}\label{labelname}%
\label{fig:webArchiveEx}
\end{center}
\end{figure}

There are several ways to search for the original post of a screenshot shared on social media. The first is using a query string in a search engine such as Google. Second, searching through fact-checking websites to see if the screenshot has been debunked. Lastly, searching web archives for the existence of a screenshot's original post (Figure \ref{fig:webArchiveEx}).\protect\footnotemark\footnotetext{\url{https://archive.ph/h96jD}} This image depicts a screenshot of a web archive page containing a Twitter post authored by @unusual\_whales. The post by @unusual\_whales contains a screenshot of two Twitter posts, the first by @OfficialLoganK and the second by @elonmusk. If the original post is found the screenshot may be real, if it is not found by any of these methods a probability of the screenshot being misattributed must be determined. The above methods are time-consuming and would benefit from the process being automated. %

\section{Related Work}
Parikh et al. \cite{ParikhFramework} developed a framework, with 83.33\% accuracy, to detect tampered and impersonated tweets on digital platforms. The framework can be described in several steps. First, a user uploads a screen capture to a server. Then the Name, Username, Tweet Text, and timestamp are extracted. The extracted data is verified and a validation model using Twitter API keys is utilized. The Twitter database is queried and an outcome of success or failure is applied. A comparison analysis is performed and used to determine if a tweet has been tampered with or not. A dataset was collected using Twitter’s public API (no longer public, as of February 2023 \cite{TwitterAPI}) consisting of 100 screenshots. This work utilizes Twitter’s API, which is currently no longer public. One limitation of this work is the inability to determine if a screenshot of a tweet has been edited if the tweet is no longer live on Twitter.

Hodges, Chaiet, and Gupta \cite{hodges2021forensic} studied the identification of memes in social media, including cross-platform propagation.  They developed tools to classify the source of social media platforms in the images, but did not focus on verifying attribution.  

Abdali et al. \cite{AbdaliWebsiteScreenshots} proposes the use of visual cues contained in a website to detect misinformation. Using a semi-surpervised classification technique, referred to as VizFake, an  $F_1$ score of 85\% on a dataset of 50,000 news article screenshots is achieved using less than 5\% of the labels. The authors argue that the method of using the article’s text to identify misinformation has several drawbacks and thus the process is benefited by using domain-level features and visual cues.

\begin{figure}
\begin{center}
\includegraphics[scale=0.75]{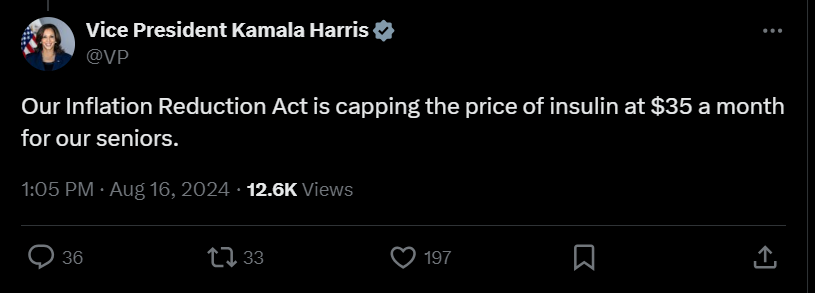}%
\captionof{figure}[h!]{Status: Posted on Twitter.}\label{labelname}%
\label{fig:Status}
\end{center}
\end{figure}

\begin{figure}
\begin{center}
\includegraphics[scale=0.2]{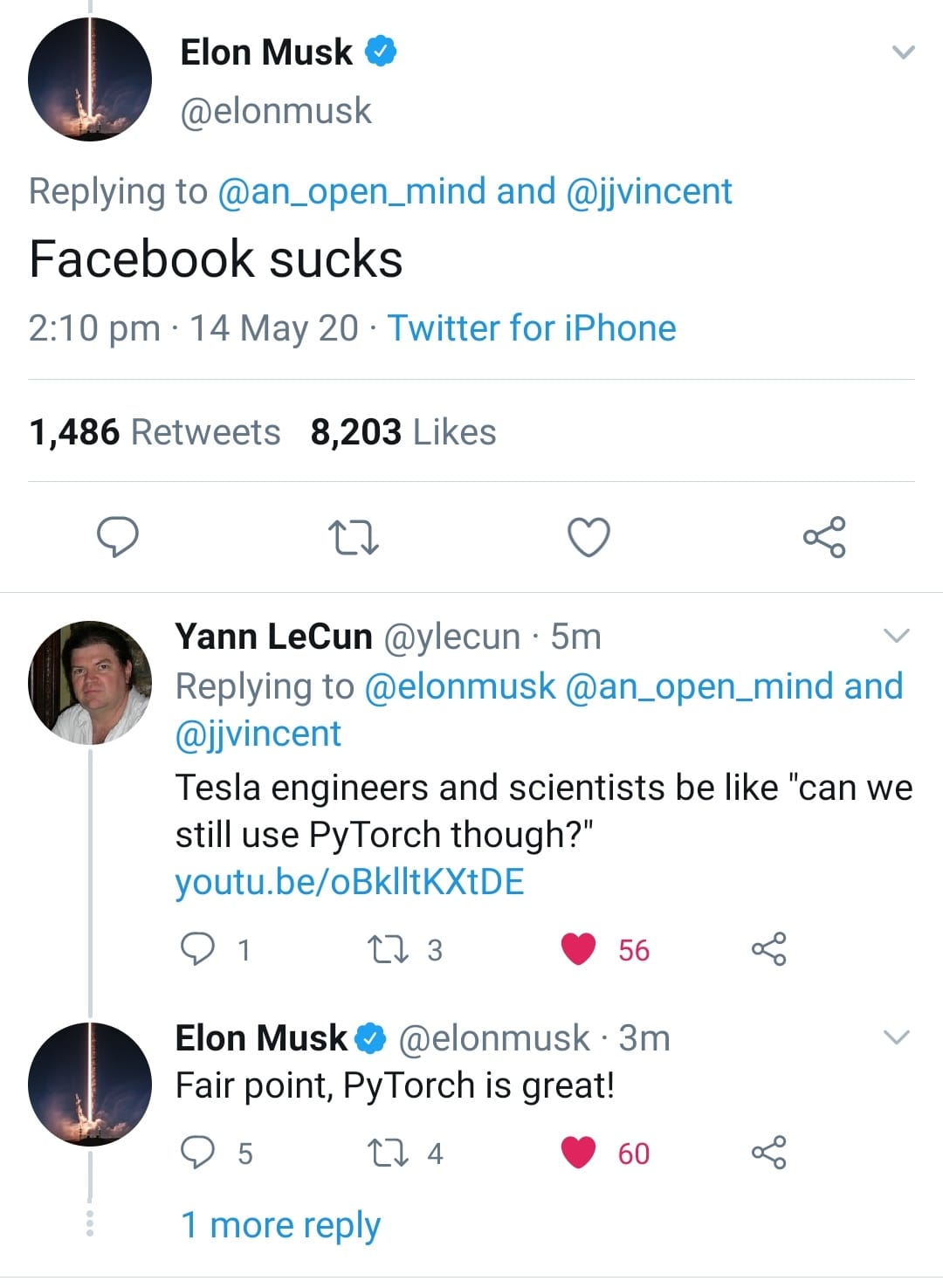}%
\captionof{figure}{Thread: Screenshot of Twitter post containing a thread.}\label{labelname}%
\label{fig:Thread}
\end{center}
\end{figure}

Zaki et al. \cite{Zaki2023ExtractingIF} is working to automate the process of screenshot validation. To do this Zaki is annotating a dataset containing screenshot images collected primarily from Twitter. Zaki manually searches for their original post on the live web or in web archives. When attempting to find the original post of a screenshot, the Tweet text, Twitter handle, and timestamp is extracted so that an effective search in the web archives can be made. To extract these elements Optical Character Recognition (OCR) is performed using {Python-tesseract\protect\footnotemark}\label{labelname}\footnotetext{\url{https://pypi.org/project/pytesseract/}} and the date is then extracted using the {datefinder\protect\footnotemark}\label{labelname}\footnotetext{\url{https://pypi.org/project/datefinder/}} module. Zaki has also identified several labels to describe Twitter posts that can be applied to other platforms. These labels (Table \ref{tab:labels}) describe the structure of a post that we can see in either an original post or a screenshot \cite{Zaki-blog}. The most simple label is called a Status\protect\footnotemark\footnotetext{\url{https://x.com/bennyjohnson/status/1797411023881654627}} and refers to a single post (Figure \ref{fig:Status}). A combination of multiple statuses connected is called a thread (Figure \ref{fig:Thread}).\protect\footnotemark\footnotetext{\url{https://x.com/elonmusk/status/1260852444818362369}} These posts generally include phrases such as "replying to" or are linked together to show the order of conversation. Co-tweets (Figure \ref{fig:denyEngagement}) specifically refers to a Twitter post with another post nested within. The final label is a cropped snapshot (Figure \ref{fig:concatenatedScreenshot}), also described as concatenated posts. These posts are created when a user combines multiple posts, this is an example of extending the functionality of a platform.

\begin{table}[h!]
    \centering
    \begin{tabular}{|c|c|}
    \hline
        \textbf{Post Type} & \textbf{Structural Features}\\
        \hline
        Status & Single Tweet\\
        \hline
        Reply & Tweet responding to another Tweet\\
        \hline
        Co-Tweet & Nested, Retweet\\
        \hline
        Cropped Snapshot & Multiple Tweets pasted together\\
    \hline
    \end{tabular}
    \vspace*{3mm}
    \caption{Labels describing the structural features of social media posts \protect\cite{Zaki2023ExtractingIF}.}
    \label{tab:labels}
\end{table}

Bradford and Nelson \cite{Bradford2022DidTR}, are developing methods of searching the web for evidence of a tweet's existence. To do this a modular system called \textit{SSAuth} is being implemented. \textit{SSAuth}, automatically searches not only the web, but {Snopes.com\protect\footnotemark}\label{labelname}\footnotetext{\url{https://www.snopes.com/}} and {Reuters.com,\protect\footnotemark}\label{labelname}\footnotetext{\url{https://www.reuters.com/}} fact-checking websites, for the text content of a tweet. He also is developing tools to search {Politwoops,\protect\footnotemark}\label{labelname}\footnotetext{\url{https://projects.propublica.org/politwoops}} a website that saves deleted political tweets. Bradford built \textit{SSAuth} using a Python script that passes the content of a tweet using command line arguments. Bradford has discovered that the most efficient queries only use the first 50 characters of a tweet.

Nwala et al. describe a vocabulary to label the internal structure of web resources \cite{Nwala2019UsingMI}. This vocabulary is based on an acronym system describing the number of posts and authors in a web resource (Table \ref{tab:internalCat}). These categories are used to describe the ways a resource naturally occurs in a system, not necessarily what we see in a screenshot. Because of this we can only apply these categories to some of the labels observed by Zaki, for example, a status would occur in the system as \(P_1A_1\), a thread and co-tweet could occur either as \(P_nA_1\) or \(P_nA_n\), but a cropped-snapshot would never occur naturally on the platform since it is an extension of its functionality.

\begin{table}[h!]
    \centering
    \begin{tabular}{|c|c|c|}
    \hline
        \textbf{Acronymn} & \textbf{Post Count} & \textbf{Author Count}\\
        \hline
        \(P_1A_1\) & Single (1) & Single (1)\\
        \hline
        \(P_1A_n\) & Single (1) & Multiple (n)\\
        \hline
        \(P_nA_1\) & Multiple (n) & Single (1)\\
        \hline
        \(P_nA_n\) & Multiple (n) & Single (1)\\
    \hline
    \end{tabular}
    \vspace*{3mm}
    \caption{Acronymn classification proposed by Alexander Nwala to categorize internal structure of web resources.}
    \label{tab:internalCat}
\end{table}

\section{Methodology}
To automatically search for original social media posts, it is important to be able to identify from which platform a screenshot originated. It is quite easy for humans to identify the platform due to the visual cues and prior knowledge of what platforms look like, but this process is difficult to automate. Besides differentiating between platforms, the number of posts within the screenshot needs to be identified and the metadata extracted from the image to make effective search queries for the original post to determine correct attribution. 

\subsection{Extracting Metadata}
\begin{figure}[h!]
\centering
\fbox{
\includegraphics[width=.46\linewidth]{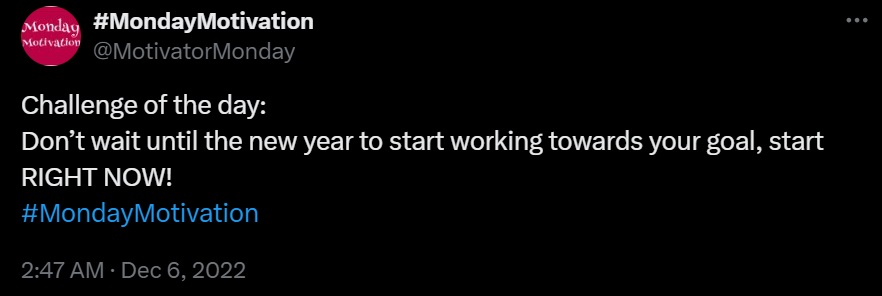}
}
\fbox{
\includegraphics[width=.46\linewidth]{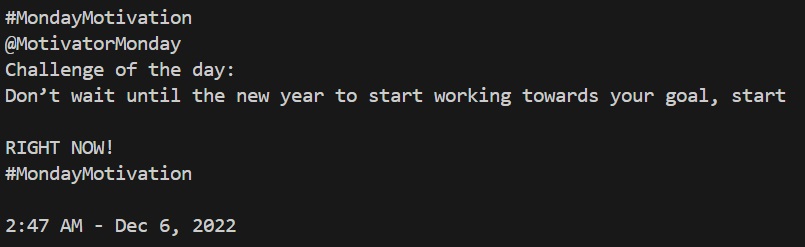}
}
\caption{
\textbf{Post compared to OCR:}
\textnormal{Side-by-side comparison of a Twitter post and the post after OCR is performed.}}
\label{fig:ocrCompare}
\end{figure}

\begin{figure}[h!]
\begin{center}
\includegraphics[scale=0.75]{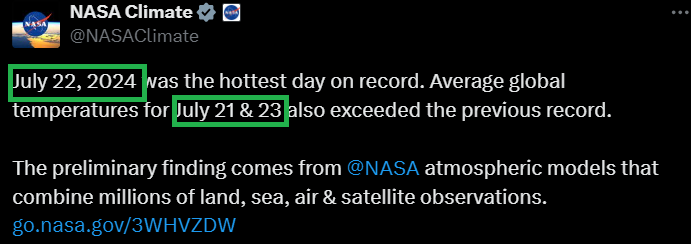}%
\captionof{figure}{Screenshot of Twitter post with inessential dates (green boxes added for emphasis, not original to post).}\label{labelname}%
\label{fig:dates}
\end{center}
\end{figure}

Before metadata could be extracted Optical Character Recognition (OCR) was performed on the screenshot using Python-tesseract (Figure \ref{fig:ocrCompare}). When an image was passed to the OCR module, a text string containing all the characters contained within the image was returned. To extract the timestamp the Python datefinder module was used to identify and reformat dates. To differentiate between valid timestamps and inessential dates (Figure \ref{fig:dates})\protect\footnotemark\footnotetext{\url{https://x.com/NASAClimate/status/1817939517094986097}} contained in the text body, the Python {RegEx\protect\footnotemark}\label{labelname}\footnotetext{\url{https://pypi.org/project/regex/}} module was utilized to create a pattern that would identify the possible formats a date could occur in specifically for Twitter. The string returned by OCR was parsed into individual lines and any line containing a date that matched the predetermined pattern was identified. Each line containing a date was then searched for any words that should not be present in the same line as the meaningful date and if found were eliminated. A text file forked from the GitHub repository, google-10000-english \cite{github20k}, was used to identify words that should not be contained in a line with a meaningful date. This isolated the meaningful dates from the inessential ones present within the text body. A similar logic to the date module was applied to differentiate authors from users mentioned in the body of a post. Authors were identified by an address sign (@) followed by 4--15 characters. Any author that was embedded in a line containing a common word was eliminated as a potential author for the post. All dates and authors were then returned in the order they occurred in the screenshot and used to create self-contained units within a single screenshot. 

\subsection{Categorizing Screenshots}
\begin{figure}
\begin{center}
\includegraphics[scale=0.75]{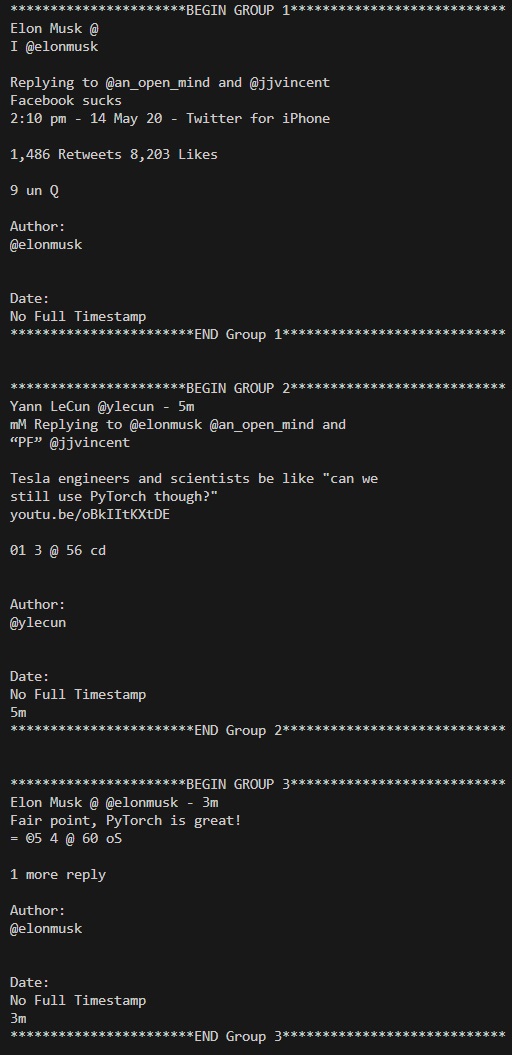}%
\captionof{figure}{Twitter posts of conversation between @elonmusk and @ylecun (Figure \ref{fig:Thread}) grouped together by metadata after OCR.}
\label{fig:successGroup}
\end{center}
\end{figure}

After all necessary metadata was extracted the posts were then categorized based on the internal structure that the original post would have likely occurred in by first grouping posts contained in the screenshot together. The number of posts was determined by the number of meaningful dates contained in the screenshot and the authors were determined by the number of meaningful usernames found in the screenshot. The number of posts and authors were used to determine the internal structure as described by Nwala et al. \cite{Nwala2019UsingMI}. Posts were separated by the location of authors, for example in the screenshot of the thread conversation between @elonmusk and @ylecun (Figure \ref{fig:Thread}), the first post was identified by all of the text above the post authored by @ylecun, the second identified by the text between the two post between @elonmusk, and the final post consisting of all the text remaining (Figure \ref{fig:successGroup}).

\subsection{Building a Dataset}
To determine which platform a screenshot originated from a collection of images from different platforms needed to be collected so that we could later build an accurate model. We developed a modular scraper system using the headless browser {Puppeteer\protect\footnotemark}\label{labelname}\footnotetext{\url{https://pptr.dev/}} to take screenshots of posts on four platforms: Twitter, Instagram, Truth Social, and Facebook. The scraper system was passed a list of URLs obtained from four separate publicly available datasets via command line arguments (Table \ref{tab:datasets}). If a full URL was not available in a dataset, the link could be built backward using the post ID. For example, when given an Instagram post ID, \textit{InstagramID}, the complete URL was built using the format https://www.instagram.com/p/\textit{InstagramID}/. The URLs were extracted from files within each dataset and saved in a .txt file (Figure \ref{fig:txtFile}). Screenshots were taken of each URL in light mode, dark mode, web version, and mobile version. If needed, screenshots were cropped using {ImageMagick.\protect\footnotemark}\label{labelname}\footnotetext{\url{https://imagemagick.org/}} Pre-built datasets were also searched for to increase the size of the dataset. We found one dataset created by Bot Sentinel \cite{BotSentinel-Dataset} that contains 1,363 screenshots of tweets that harassed Kamala Harris. These images were cropped using ImageMagick and added to the total of the Twitter mobile light category.

\begin{table}[h!]
    \centering
    \begin{tabular}{|c|c|c|}
    \hline
        \textbf{Platform} & \textbf{URL} & \textbf{Citation}\\
        \hline
        Facebook & https://www.kaggle.com/datasets/mchirico/cheltenham-s-facebook-group & \cite{Facebook-Dataset}\\
        \hline
        Instagram & https://www.kaggle.com/datasets/shmalex/instagram-images & \cite{Instagram-Dataset}\\
        \hline
        Truth Social & https://zenodo.org/records/7531625 & \cite{TruthSocial-Dataset}\\
        \hline
        Twitter & http://nlp.uned.es/replab2013/ & \cite{Twitter-Dataset}\\
    \hline
    \end{tabular}
    \vspace*{3mm}
    \caption{Dataset links from which the URLs were obtained and used to take screenshots of each platform.}
    \label{tab:datasets}
\end{table}

\begin{figure}
\noindent
\texttt{https://twitter.com/BMW\_LifeMorals/status/241301682477232128
https://twitter.com/BMW\_LifeMorals/status/253567721424445441
https://twitter.com/BMW\_LifeMorals/status/257181645101211649
https://twitter.com/BMW\_LifeMorals/status/261522710146977792
https://twitter.com/BMW\_LifeMorals/status/261995784894029824
https://twitter.com/BMW\_LifeMorals/status/269467393900830721
https://twitter.com/BMW\_LifeMorals/status/272380339362623488
https://twitter.com/BMW\_LifeMorals/status/272402989015244801
https://twitter.com/BMW\_LifeMorals/status/272873624023732224
}
\captionof{figure}{.txt file containing Twitter URLs.}
\label{fig:txtFile}
\end{figure}

\section{Results and Discussion}

To evaluate the categorization of Twitter posts by internal structure, 75 screenshots that were shared on Twitter were collected and annotated by hand. Each screenshot was categorized by its internal structure. The categorization script was then tested on these 75 screenshots and a precision, recall, and $F_1$ score was calculated for each internal category as well as the mean scores (Table \ref{tab:PreRecF1}). Note the deficiency in the number of screenshots in the internal category \(P_nA_1\), this category had significantly fewer images than the other three and thus there an increase in the number of images would drastically change the evaluation metrics for this category. The internal category of \(P_1A_1\) performed the best, as would be expected since it is the least complicated form of Twitter Post. The most common reason a post was categorized incorrectly was an incorrect extraction of text from the image when OCR was performed.

\begin{table}[h!]
    \centering
    \begin{tabular}{|c|c|c|c|}
    \hline
        \textbf{Category} & \textbf{Precision} & \textbf{Recall} & \textbf{$F_1$}\\
        \hline
        \(P_nA_n\) (k=18) & 0.93 & 0.72 & 0.81\\
        \hline
        \(P_nA_1\) (k=4) & 0.50 & 0.75 & 0.60\\
        \hline
        \(P_1A_1\) (k=53) & 0.95 & 0.98 & 0.96\\
        \hline
        Overall (k=75) & 0.79 & 0.82 & 0.80\\
    \hline
    \end{tabular}
    \vspace*{3mm}
    \caption{Precision, recall, and $F_1$ score for classification of images ($k=75$) by internal structure.}
    \label{tab:PreRecF1}
\end{table}

\begin{figure}
\begin{center}
\includegraphics[scale=0.75]{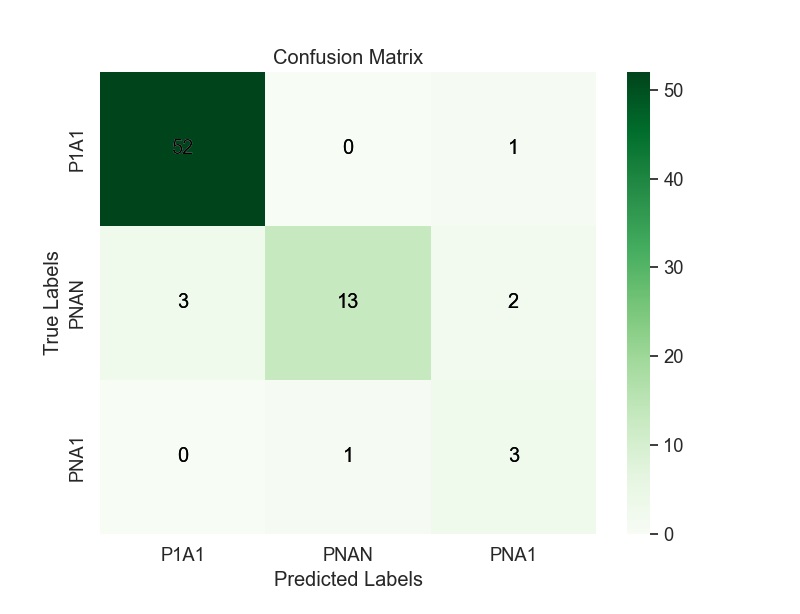}%
\captionof{figure}{Confusion matrix for number of images ($k=75$) placed into the correct internal category.}
\label{fig:confusionMatrix}
\end{center}
\end{figure}

A confusion matrix (Figure \ref{fig:confusionMatrix}) was then generated to demonstrate the number of images that were categorized correctly and incorrectly by class. The matrix shows the true labels, what the actual structure of the post was, and how many times they were classified correctly. The predicted labels, the structure the classifier assigned to a post, show how many times the image were correctly and incorrectly classified by our script.

\begin{figure}
\begin{center}
\includegraphics[scale=0.75]{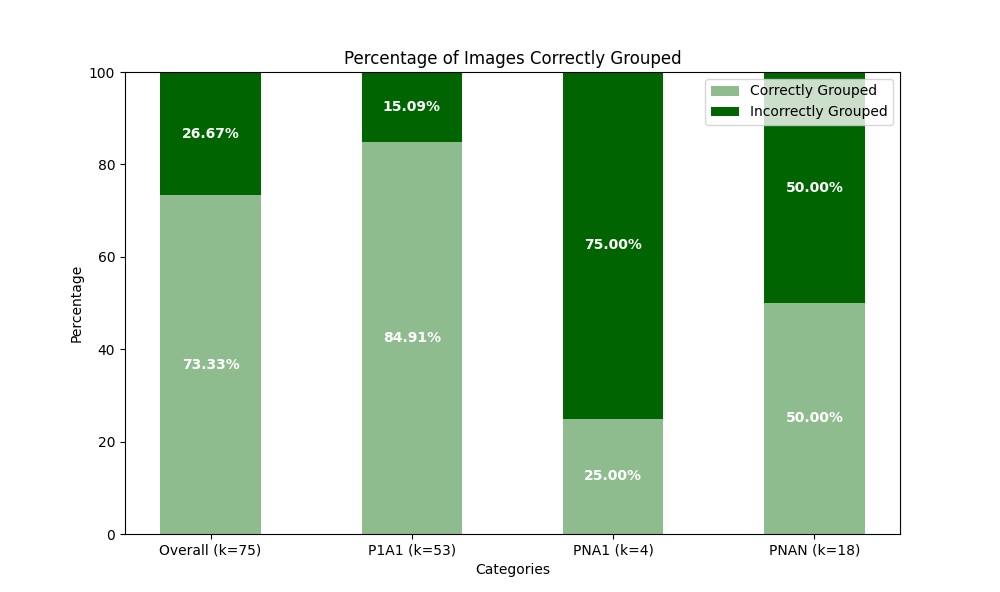}%
\captionof{figure}{Percentage of Images ($k=75$) grouped by metadata broken down by internal structure.}
\label{fig:barChart}
\end{center}
\end{figure}

A stacked bar chart (Figure \ref{fig:barChart}) was generated to demonstrate the percentage of images grouped by their metadata correctly and incorrectly. This means that for every post within a screenshot, the author, timestamp, and text body were grouped correctly. The bar chart displays the percentage of all images ($k=75$) grouped correctly and incorrectly alongside the percentage of metadata grouped correctly for images belonging to each structure. Again, due to the deficiency in images within the internal structure \(P_nA_1\) it is difficult to draw conclusions based on this category. However, we can see about 73\% of the images were grouped by their metadata correctly. A difficulty presented with the category \(P_nA_n\) was concatenated images. Often a concatenated image would not contain the necessary metadata to group the images. A second complication was OCR not extracting text from an image accurately. This would distort the posts and make them difficult to group.

To eventually build a model to effectively differentiate between various platforms a large dataset first needed to be created for training and testing purposes. Images were collected from Facebook, Instagram, Truth Social, and Twitter in light mode, dark mode, web version, and mobile version. Since four platforms were being focused on there were four categories for each platform, 16 social media scrapers were built to capture screenshots of the posts. Each scraper was passed the .txt file of URLs as an argument and took screenshots of all non-broken URLs. Overall we were able to collect about 16,000 images (Table \ref{tab:kImages}). 

Several barriers needed to be crossed when taking screenshots of social media posts. For example, when attempting to only capture the contents of a post instead of the entire screen an HTML element needed to be identified to mark the boundaries of the post. However, the Facebook web version does not contain a meaningful element of the isolated post that can be used to do this. To overcome this a screenshot was taken by the scraper of the full web page and ImageMagick was used to crop all the images collected. A second issue that needed to be overcome was the potential of the platform to recognize the scraper as a bot. Several methods were used to overcome this with varying success. A sleep function was added to the scrapers to pause for a random amount of time between two and twelve seconds. A second delay was added to pause the scrapers for five minutes for every 100th URL. This allowed more images to be taken at a time without significant delay, however, in the case of Instagram only several hundred images could be taken without having to pause for an extended period due to the platform no longer loading posts. \\
\newline
\begin{table}[ht!]
\begin{center}
\begin{tabular}{|l|c|c|c|c|}\hline
\diaghead{Diag ColumnmnHead II}%
{\textbf{Web}/\textbf{Mobile}\\\textbf{Light}/\textbf{Dark}}
{\textbf{Platform}}&
\textbf{FB}&\textbf{IG}&\textbf{TS}&\textbf{T}\\     \hline
\textbf{MD}&\thead{788}&\thead{454}&\thead{1,777}&\thead{1,027}\\     \hline
\textbf{ML}&\thead{797}&\thead{524}&\thead{1,781}&\thead{2,563}\\     \hline
\textbf{WD}&\thead{989}&\thead{398}&\thead{848}&\thead{1,285}\\     \hline
\textbf{WL}&\thead{987}&\thead{823}&\thead{846}&\thead{734}\\      \hline
\textbf{Platform Total}&\thead{3,561}&\thead{2,199}&\thead{5,252}&\thead{5,609}\\     \hline
\end{tabular}
\vspace*{3mm}
\captionof{table}{Number of images collected using scrapers.}
\label{tab:kImages}
\end{center}
\end{table}

\section{Conclusions}

False information in the form of misattribution is often spread by sharing screenshots on social media. Though these screenshots help extend the functionality of a platform they often are used to persuade others to believe someone is saying something that they never actually said. Because a screenshot has no link back to the original post, it is difficult to tell if the post is being correctly attributed without searching the web or internet archives. This calls for tools to be developed to automate the search process for an original tweet. This project aims to develop tools to aid in the process of automating queries for an original post. So far this research has focused on extracting metadata from screenshots of Twitter posts, assigning a category to these screenshots based on the internal structure they would likely occur in on the Twitter platform, and grouping all posts captured in a screenshot together by their metadata. We evaluated the categorizer module of the script by performing a precision and recall test and calculating the $F_1$ score. We found that the overall precision was 0.79, the recall was 0.82, and the $F_1$ score was 0.80. To understand how well the grouping module of the script was performing, we calculated the percentage of images correctly grouped by their metadata and found it to be 73.33\%. We also collected a dataset that can aid in training and testing a model to differentiate between various social media posts. The platforms focused on in this dataset were Facebook, Instagram, Truth Social, and Twitter. So far over 16,000 images have been collected.

\section{Acknowledgements}
This work is supported in part by the National Science Foundation Research Experience for Undergraduates Site Award \#2149607. We would like to thank the graduate students in the Web Sciences and Digital Libraries Research Group at Old Dominion University, especially Tarannum Zaki, for their assistance with this work, as well as the faculty who dedicated their time and experience to advising this project. We are also very appreciative of Prof. Jennifer Golbeck and her recommendation of the dataset created by Bot Sentinel.

\newpage
\renewcommand{\refname}{References Cited}
\bibliographystyle{ieeetr}
\bibliography{refs} 

\begin{thebibliography}{10}

\bibitem{Apa-Misinformation/Disinformation}
{American Psychological Association}, ``Misinformation and disinformation.'' https://www.apa.org/topics/journalism-facts/misinformation-disinformation, 2022.

\bibitem{ParikhFramework}
S.~B. Parikh, S.~R. Khedia, and P.~K. Atrey, ``A framework to detect fake tweet images on social media,'' in {\em 2019 IEEE Fifth International Conference on Multimedia Big Data (BigMM)}, pp.~104--110, 2019.

\bibitem{TwitterAPI}
A.~Blok, ``Twitter {API} is going behind the paywall.'' https://www.cnet.com/news/social-media/twitter-api-is-going-behind-the-paywall/, 2023.

\bibitem{hodges2021forensic}
J.~A. Hodges, M.~Chaiet, and P.~Gupta, ``Forensic analysis of memetic image propagation: introducing the {SMOC BRISQUEt} method,'' {\em Proceedings of the {Association for Information Science and Technology}}, vol.~58, no.~1, pp.~196--205, 2021.

\bibitem{AbdaliWebsiteScreenshots}
S.~Abdali, R.~Gurav, S.~Menon, D.~Fonseca, N.~Entezari, N.~Shah, and E.~E. Papalexakis, ``Identifying misinformation from website screenshots,'' in {\em Vol. 15 (2021): Fifteenth International {AAAI} Conference on Web and Social Media}, pp.~2--13, 2021.

\bibitem{Zaki2023ExtractingIF}
T.~Zaki, M.~L. Nelson, and M.~C. Weigle, ``Extracting information from twitter screenshots,'' Tech. Rep. arXiv:2306.08236, 2023.

\bibitem{Zaki-blog}
T.~Zaki, ``Disinformation spread on social media through screenshot sharing: Dataset description.'' https://ws-dl.blogspot.com/2022/12/2022-12-12-disinformation-spread-on.html, 2022.

\bibitem{Bradford2022DidTR}
C.~Bradford and M.~L. Nelson, ``Did they really tweet that? {Q}uerying fact-checking sites and {P}olitwoops to determine tweet misattribution,'' Tech. Rep. arXiv:2211.09681, 2022.

\bibitem{Nwala2019UsingMI}
A.~C. Nwala, M.~C. Weigle, and M.~L. Nelson, ``Using micro-collections in social media to generate seeds for web archive collections,'' in {\em Proceedings of the 2019 {ACM/IEEE} Joint Conference on Digital Libraries {(JCDL)}}, pp.~251--260, 2019.

\bibitem{github20k}
J.~Kaufman, ``google-10000-english.'' https://github.com/first20hours/google-10000-english, 2021.

\bibitem{BotSentinel-Dataset}
{Bot Sentinel Inc.}, ``{T}witter's response to abuse and bigotry directed at vice president {K}amala {H}arris.'' https://botsentinel.com/reports/documents/kamala-harris/report-05-26-2022.pdf, 2022.

\bibitem{Facebook-Dataset}
M.~Chirico, ``Cheltenham's facebook groups.'' https://www.kaggle.com/datasets/mchirico/cheltenham-s-facebook-group/data, 2018.

\bibitem{Instagram-Dataset}
A.~Matusevski, ``Instagram images - 1,211,625 posts.'' https://www.kaggle.com/datasets/shmalex/instagram-images, 2022.

\bibitem{TruthSocial-Dataset}
P.~Gerard, N.~Botzer, and T.~Weninger, ``Truth social dataset [data set].'' https://doi.org/10.5281/zenodo.7531625, 2023.

\bibitem{Twitter-Dataset}
E.~Amig{\'o}, J.~{Carrillo de Albornoz}, I.~Chugur, A.~Corujo, J.~Gonzalo, T.~Mart{\'i}n, E.~Meij, M.~de~Rijke, and D.~Spina, ``Overview of {RepLab} 2013: Evaluating online reputation monitoring systems,'' in {\em {Proceedings of the Fourth International Conference of the CLEF initiative}}, pp.~333--352, 2013.

\end{thebibliography}
\end{document}